*Article*

# How does dissipation affect the transition from static to dynamic macroscopic friction?

**Naum I. Gershenzon [1],\*, Gust Bambakidis [2], and Thomas Skinner [2]**

[1] Physics Department & Department of Earth and Environmental Sciences, Wright State University, 3640 Colonel Glenn Highway, Dayton, OH 45435; E-Mail: naum.gershenzon@wright.edu

[2] Physics Department, Wright State University, 3640 Colonel Glenn Highway, Dayton, OH 45435; E-Mail: gust.bambakidis@wright.edu; thomas.skinner@wright.edu

**\*** Author to whom correspondence should be addressed; E-Mail: naum.gershenzon@wright.edu; Tel.: +1-937-775-2052.

External editor:

**Abstract:** Description of the transitional process from a static to a dynamic frictional regime is a fundamental problem of modern physics. Previously we developed a model based on the well-known Frenkel-Kontorova model to describe dry macroscopic friction. Here this model has been modified to include the effect of dissipation in derived relations between the kinematic and dynamic parameters of a transition process. The main (somewhat counterintuitive) result is a demonstration that the rupture (i.e. detachment front) velocity of the slip pulse which arises during the transition does not depend on friction. The only parameter (besides the elastic and plastic properties of the medium) controlling the rupture velocity is the shear to normal stress ratio. In contrast to the rupture velocity, the slip velocity does depend on friction. The model we have developed describes these processes over a wide range of rupture and slip velocities (up to 7 orders of magnitude) allowing, in particular, the consideration of seismic events ranging from regular earthquakes, with rupture velocities on the order of a few km/s, to slow slip events, with rupture velocities of a few km/day.

## 1. Introduction

The relative movement of two solids in contact is accompanied by friction, an essentially nonlinear dissipative process. It is generally accepted that friction appears due to interactions between surface asperities. The actual contact area between rough frictional surfaces of stiff materials is less (usually much less) than the nominal surface area, is proportional to the averaged normal stress, and depends on the elasticity and plasticity of the materials in contact [1, 2]. The normal stress at the tip of an asperity (i.e., at the physical contact area) is equal to the penetration hardness of the material [2]. Under static or uniform sliding conditions, friction is usually described by the frictional coefficient, i.e. the proportionality coefficient between tangential and normal stress (classical Amontons-Coulomb law). However, as shown in modern laboratory experiments, friction depends on slip, sliding rate, contact time and normal stress history (see extensive reviews by Marone [3]), Baumberger and Caroli [4] and Dieterich [5]). Recent laboratory experiments [6-9] also confirm that such a description is not sufficient when parameters describing the physical state of the system (sliding rate, stress, etc.) are not uniform in time and/or space.

Over the past 50 years, various approaches for the modeling of non-uniform frictional processes have been developed. Two types of models are the most common, i.e. mass-spring models [10-17], and rate-and-state (Dietrich-Ruina) models [18-28]. The mass-spring models of the Burridge-Knopoff type describe collective behavior and statistical features of earthquakes and reproduce major empirical laws of observed seismicity, i.e., large earthquake recurrence, the Gutenberg-Richter law, foreshock and aftershock activities, and preseismic quiescence [10, 11, 13]. In general, however, these models are not adequate to describe the dynamics of an individual event. More detailed dynamics, such as the necessary and sufficient conditions for nucleation of individual earthquake events, have instead been formulated in the framework of rate-and-state models [19-21]. Rate-and-state models have been used successfully to describe regular earthquakes [18-23], slow slip events [24, 25], and fault dynamics [e.g. 22, 23, 26]. They are capable of incorporating such phenomena as frictional dilatancy [29-31], compaction of brittle materials [29, 32] and microscopic elasticity [33]. Although these models are based on laboratory experiments and include some measurable laboratory parameters, such as the characteristic slip distance, they also include unknown parameters which can be adjusted to fit field or laboratory observations. The same is true for the mass-spring models. Ultimately, a physical model with no adjustable parameters is most desirable.

The Frenkel-Kontorova (FK) model [34] provides a promising point of departure for a more predictive model. It has been widely used to describe micro- and nano-scopic friction (e.g. [35, 36] and references therein). Recently, we have developed a FK-type model which describes *macroscopic* friction. The advantages of this model are: 1) it is an intrinsically dynamical model, rooted in the Newtonian equations of motions; 2) parameters used in the model have explicit and unambiguous physical correlates; 3) it describes frictional processes over a wide range of conditions, from very fast processes such as regular earthquakes down to very slow processes such as creep, silent, and slow earthquakes [37-39]. The observed nonlinear dynamics of frictional processes is incorporated in the standard linear mass-spring models by introducing ad hoc nonlinear relations between various model parameters (e.g., the introduction of a nonlinear spring constant or a nonlinear relation between friction and slip velocity). By contrast, the FK model is inherently nonlinear.

The motivation for using this model to describe dry *macroscopic* friction derives from the similarity between plasticity and dry friction, both on laboratory and geophysical scales. Sporadic local motions of the Earth's crust along faults, occurring due to earthquakes and various creeping events, are similar to the processes of plastic deformation in crystals resulting from movement of edge dislocations by the localized shift of crystalline planes. Of particular relevance is the fact that the external stress initiating plasticity is only a small fraction of the stress necessary for the uniform relative displacement of planes of crystal atoms. Similarly, laboratory friction experiments have shown that the critical shear force needed to initiate a macroscopic slip pulse between frictional surfaces is usually much less than that predicted by theory [7].

Motivated by these similarities, we have proposed a novel model [37, 38] in which sliding occurs in much the same way as in plasticity, i.e., due to movement of a certain type of defect, a "macroscopic dislocation," which requires much less shear stress than uniform displacement of frictional surfaces. A dislocation is a static configuration of accumulated stress between two surfaces due to the elastic shift of asperities on one surface relative to the other. The dislocation is confined to a specific region of the surface but can be displaced along the surface. As we will see, the sliding motion of the two surfaces occurs due to the movement or propagation of dislocations, somewhat analogous to caterpillar motion. A macroscopic slip is, in fact, the result of a multitude of dislocations propagating along the surface.

In the continuum limit, our model is described by the sine-Gordon (SG) equation, one of the fully integrable nonlinear equations of mathematical physics. This equation has been thoroughly investigated due to its exceptional importance and universality [40-43]. The mathematical apparatus which has been developed is fully applicable to the problems considered here. In the framework of our model, all variables, whether or not directly measurable, are connected by transcendental analytical relations, allowing a clear analysis of dependencies, e.g., rupture velocity as a function of accumulated shear stress [37, 38]. Algebraic relations have been obtained between kinematic parameters (such as slip velocity and rupture velocities) and dynamic parameters (such as shear stress, normal stress, and stress drop). However these formulae (analytical solutions) have been derived neglecting friction. Here, we introduce a dissipative term into the SG equation, which requires a numerical solution of the problem. We show that some of our previous results, such as the relation between rupture and shear stress, remain valid under the influence of dissipative processes, but some of them, such as the relation between slip velocity and shear stress, need to be modified.

In the next section we describe the basics of the model, followed by the Results, Discussion, and Conclusion.

## 2. Model

An overview of the model is provided to establish the context for the results which follow. A more detailed description may be found in our previous articles [37, 38].

2.1 *Model derivation*

Asperities on each frictional surface are idealized as uniform sinusoidal surfaces, illustrated in Figure 1. We will consider asperities on one of the frictional surfaces as forming a linear chain of balls of mass *M*, each ball interacting with its nearest neighbors via spring forces of stiffness $K_b$. These provide the forces of elastic deformation for the shift of an asperity from its equilibrium position. The asperities on the opposite frictional surface are regarded as forming a rigid substrate which interacts with the masses *M* via a sinusoidal restoring force. The physical correlate of this force is the horizontal component of the normal force exerted by the lower surface on upper plate asperities displaced as in figure 1a. Application of this model to describe the slip dynamics yields the FK model, where we have also included an explicit frictional force $f_i$ on the $i^{th}$ asperity:

$$M \frac{\partial^2 u_i}{\partial t^2} - K_b(u_{i+1} - 2u_i + u_{i-1}) + F_d \sin \frac{2\pi}{b} u_i = F(x,t) - f_i(x,t, \frac{\partial u_i}{\partial t}), \quad (1)$$

where $u_i$ is the shift of ball (asperity) *i* relative to its equilibrium position, *b* is a typical distance between asperities, *t* is time, $F_d$ is the amplitude of the periodic restoring force and *F* is the external (or driving) force. In this model, only the interaction between nearest-neighbor asperities is considered. This approximation is motivated by the fact that a disturbance of the stress field around an asperity decreases with distance as $r^{-3}$, thus the interaction between nearest neighbor asperities is at least 8 times larger than the interaction between the next closest asperities twice as far away.

**Figure 1.** Schematic of asperity contact (**a**) and chain of masses interacting via elastic springs and placed in a periodic potential (substrate) (**b**). The balls represent asperities. The sine-shaped surface is the opposite plate.

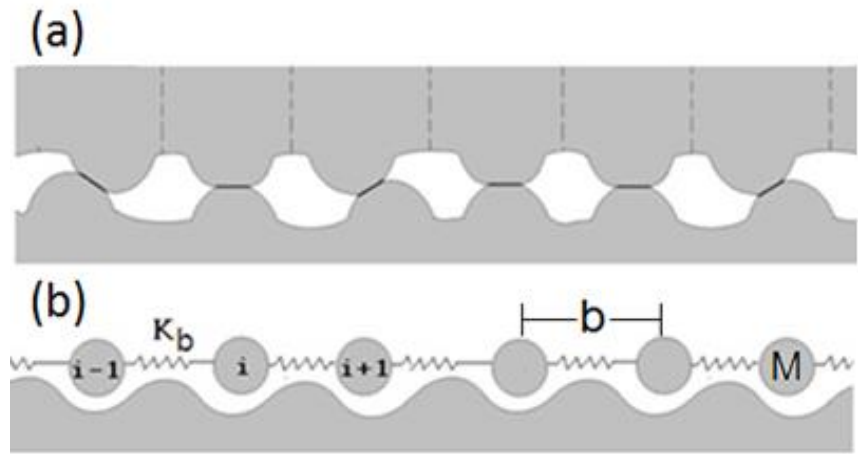

Following the same procedure used to describe plasticity [42, 44, 45], we express the coefficients of equation (1) through the parameters of the material and the frictional surfaces. For a volume density $\rho$ we find: $M = \rho b^3$, $K_b = \frac{2\mu b}{(1-\nu)}$, $F_d = \frac{\mu b^2}{2\pi}$, where $\mu$ is the shear modulus and $\nu$ the Poisson ratio. Then equation (1) can be written in the continuous form:

$$\frac{\partial^2 (2\pi u/b)}{\partial (tc/b)^2} - \frac{\partial^2 (2\pi u/b)}{\partial (x/b)^2} + A^2 \sin(\frac{2\pi u}{b}) = (F-f) \frac{2\pi A^2}{\mu b^2}, \quad (2)$$

where $c^2 = 2\mu/(\rho(1-\nu)) \equiv c_l^2(1-2\nu)/(1-\nu)^2$ and $c_l$ is the longitudinal acoustic velocity (or P wave velocity). The dimensionless parameter $A$ is equal to $[(1-\nu)/2]^{1/2}$. Note that in deriving the FK model to describe plasticity in crystals, $A^2$ is essentially the ratio of the amplitude of two forces: one is the force amplitude between an atom and the substrate layer and the other is the force amplitude between neighboring atoms at the top layer. To describe the respective coefficients for the situation where slip occurs between two external surfaces in contact, we will use equation (2) with one significant change: we shall treat the parameter $A$ phenomenologically, using the result for a crystal as a guide. So we assume that $A$ likewise depends on the ratio of two relevant forces. The force amplitude experienced by an asperity due to neighboring asperities along the slip direction is exactly the same as it was for the case of plasticity. But the force amplitude between asperity and substrate is different and depends on the normal stress $\Sigma_N$. Indeed, when $\Sigma_N = 0$ the force is zero, since there is no interaction between asperities and a substrate. On the other hand, when $\Sigma_N$ reaches the penetration hardness $\sigma_p$, the interface between the two blocks disappears and the corresponding force amplitude is essentially the same as in the case of plasticity. So we regard $A$ as a function of the ratio of $\Sigma_N$ to $\sigma_p$: $A = f(\Sigma_N/\sigma_p)$. The simplest choice is $A = \frac{((1-\nu)/2)^{1/2}\Sigma_N}{\sigma_p} \approx \frac{\Sigma_N}{\sigma_p}$. Thus, the coefficient $A$ reflects how deeply the asperities from two opposing surfaces interpenetrate and it is the ratio between actual and nominal contact areas [1, 2].

Equation (2) in dimensionless form is

$$\frac{\partial^2 u}{\partial t^2} - \frac{\partial^2 u}{\partial x^2} + \sin(u) = F - f, \qquad (3)$$

where $u$, $x$ and $t$ are now in units of $b/(2\pi)$, $b/A$ and $b/(cA)$, respectively, $F$ and $f$ are the external force and frictional force per unit area in units of $\mu A/(2\pi)$, and the derivatives $\varepsilon = \partial u/\partial x$ and $w = \partial u/\partial t$ are interpreted as the dimensionless strain and the dimensionless slip in units of $A/(2\pi)$ and $cA/\pi$, respectively. Since the driving force is the tangential stress, we set it equal to the *xz* component of the 3D stress tensor, $\sigma_s = 2\mu\varepsilon$. Thus the dimensionless stress is measured in units $\mu A/\pi$.

## 2.2 *Uniform sliding motion*

Equation (3) in the absence of external and frictional forces is the well-known SG equation. Let us consider the existence of a classical wave solution traveling to the right with wave velocity $U$ (in units of $c$) and wave number $k$ (in units of $A/b$), in which $u$ is a function of $x - Ut$. Define $\xi = x - Ut$ and $\theta = k\xi$. The basic solutions are phonons, breathers, and kinks (a particular class of solitons) [41, 42]. Only the soliton solutions represent the wave propagation necessary to model frictional dynamics. These solutions are characterized by $|U|<1$. Integrating equation (3) for the case $F - f = 0$ with these constraints gives a solution for $u$ and its derivatives in terms of the elliptic Jacobi functions, *cn* and *dn* [41]:

$$\begin{aligned} u &= \arcsin[\pm cn(\beta\xi)], \\ \sigma_s &= 2\beta\, dn(\beta\xi), \\ w &= U\sigma_s, \end{aligned} \qquad (4)$$

$$k = 2\pi N = \frac{\pi\beta}{K(m)},$$

where $K(m)$ is the complete elliptic integral of the first kind of modulus $m$ ($0 \leq m \leq 1$), $\beta = [m(1-U^2)]^{-1/2}$, and $N$ is the density of kinks in units of $A/b$. Solution (4) describes an infinite sequence of interacting kinks (solitons) of one sign, which are periodic in space and time. In the context of our model, a soliton is a dislocation.

It is also useful to introduce three dimensionless variables averaged over an oscillation period

$$W = \int \frac{w d\theta}{2\pi} = \frac{UN}{2\pi}, \Sigma_S = \int \frac{\sigma_s d\theta}{2\pi} = k, E = \int \frac{\varepsilon d\theta}{2\pi}. \tag{5}$$

These variables correspond to the measurable parameters of slip velocity, stress and strain. The parameters of a dislocation (amplitude of stress $\sigma_s^0$ and strain $\varepsilon^0$ associated with the presence of the dislocation) are: $\sigma_s^0 = \mu A/\pi$, $\varepsilon^0 = A/(2\pi)$.

Let us describe a scenario of frictional processes in terms of solutions (4) and (5). The macroscopic dislocations are nucleated on the surface by an applied shear stress in the presence of asperities. As in crystals, the mobility of a macroscopic dislocation over the frictional surface is much larger than the mobility of the whole surface, since the displacement of a dislocation (a pre-stressed area) requires less external stress. So the relative sliding of two bodies occurs due to movement of dislocations. The passage of a dislocation through a particular point on the sliding surface shifts the contacted bodies locally by a typical distance $b$. Such a dislocation may propagate with any velocity $U$ ranging from 0 to $c$, and the average velocity of sliding, i.e. the observable slip rate $W$, is proportional to the dislocation velocity $U$ and dislocation density $N$. The parameters of a dislocation (stress amplitude and pulse width) are entirely defined by the material parameters and the normal stress and do not depend on process parameters such as dislocation density and slip rate. The characteristic width of a dislocation is $D \approx 2\pi b/A$. Usually, the dislocation width is much larger than the typical distance between asperities ranging from $10^2$ to $10^4$ of the asperity size. Figure 2 schematically illustrates the frictional processes via movement of macroscopic dislocations.

> **Figure 2.** Schematics of frictional sliding of two bodies via movement of a macroscopic dislocation. The slip velocity field, *W*, of the upper sample relative to the lower sample, illustrated by the arrows, is spatially uniform at large scales (panel (**a**)). *W* is the slip velocity *w* averaged over an oscillation period in time. By contrast, the *w* velocity field is spatially non-uniform at smaller scales, i.e., on a scale comparable with a dislocation width (panel (**b**)). The spatial and temporal averages of *w* in (b) give the uniform field in (a). The dislocation size is usually much larger than the typical size of an asperity. The relative movement on the frictional surfaces at even smaller scales (asperity size) could be larger than average if it is at the center or peak amplitude of the dislocation (panel (**c**)) or smaller than average if it is in between two dislocations (panel (**d**)). The values in the panels reflect experiments described in [6-9].

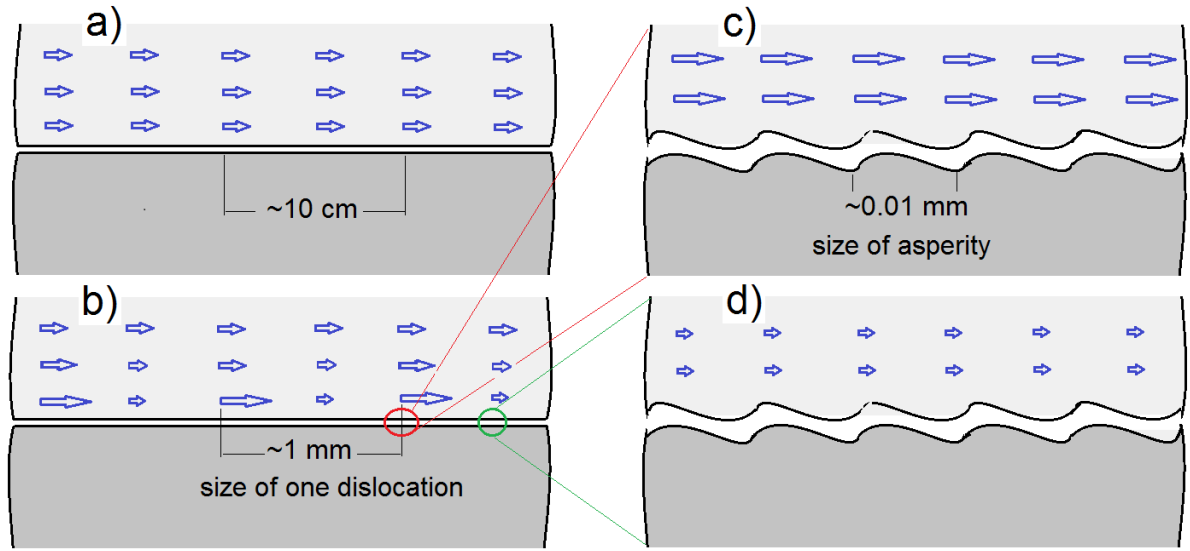

## 2.3 Non-uniform sliding motion

Solutions (4) and (5) can be used to describe the frictional process in the macroscopically uniform case. However, in the non-uniform case the macroscopic parameters such as slip velocity and stress are functions of time and space. In typical cases, the frictional surface includes many dislocations. A numerical solution for the exact positions of a multitude of microscopic dislocations is too computationally demanding to be practicable. Moreover, they cannot be measured and are therefore of little value *except* as they contribute on the average to the measured macroscopic dislocation. We therefore apply a technique which simplifies equation (3) at the expense of losing non-essential details of the process such as the exact position of each dislocation. Witham [46] developed a procedure for constructing a system of equations describing the dynamics of averaged variables (i.e., the variables in equation 5) that derive from the original dynamical equations (i.e., equation (3)). For strict applicability, the average values should vary slowly in time and space. The technique has also been found to be more generally valid in some specific cases that fail to satisfy the slowly varying constraint. For the present application, the variables of interest can all be expressed in terms of the two independent variables $U$ and $m$. Applying this procedure to the homogeneous ($F - f = 0$) SG equation yields the system of coupled equations [45]:

$$U_t \frac{\eta}{U^2 - 1} + m_t \frac{U}{2m} + U_x \frac{U\eta}{U^2 - 1} + m_x \frac{1}{2m} = 0,$$
$$U_t \frac{U}{U^2 - 1} + m_t \frac{\eta}{2mm_1} + U_x \frac{1}{U^2 - 1} + m_x \frac{U\eta}{2mm_1} = 0, \quad (6)$$

where $\eta = \frac{E}{K}$, $m_1 = 1 - m$, and $E(m)$ is the complete elliptic integral of the second kind. The system (6) is fully integrable, i.e., the solutions may be expressed in terms of analytical functions. General solutions are obtained in the references [43,45]. However, system (6) does not include friction. Whitham [46] also described a formal procedure for including a dissipative term. Applying this procedure we obtain the Whitham equations for equation (3):

$$-\frac{4}{\pi} \frac{K}{[m(1-U^2)]^{1/2}} [U_t \frac{\eta}{U^2-1} + m_t \frac{U}{2m} + U_x \frac{U\eta}{U^2-1} + m_x \frac{1}{2m}] = D(m,U),$$

$$U_t \frac{U}{U^2-1} + m_t \frac{\eta}{2mm_1} + U_x \frac{1}{U^2-1} + m_x \frac{U\eta}{2mm_1} = 0,$$

(7)

where $D(m,U) = \frac{1}{2\pi} \int [F - f(u, u_t)] d\theta$. This system of equations does not have analytical solutions (in contrast to system (6)) and must be solved numerically.

## 3. Results

We first we describe the results derived from the analytical solutions of equations (6). We then compute solutions of equations (7), which includes a dissipative term, and compare the results in order to characterize the effects of friction. Finally, we consider how the initial stress distribution affects the solution. Mathematica was used to obtain numerical solutions of (7). The initial value $U(t=0,x) = 0$ throughout all the calculations. Initial values for *m* are obtained through the definition of $\Sigma_s$ from equations (4) and (5).

*3.1. Analytical solution (no dissipation)*

To model the transition from the static to the dynamic regime of the frictional process, we consider the following idealized problem. Suppose the point $x = 0$ divides the areas of stressed ($x < 0$) and unstressed ($x > 0$) material, modeled as a step-function $\Sigma_s(t=0, x<0) = \Sigma^-$, $\Sigma_s(t=0, x>0) = \Sigma^+ = 0$, $U(t=0,x) = 0$ (solid line in Figure 3). With these initial conditions, we have $\Sigma^\pm = \frac{\pi}{\sqrt{m^\pm}K(m^\pm)}$. This setting models a stress accumulation in front of an obstacle placed at position $x = 0$ (e.g., a large asperity). The effect of alternative initial shear stress profiles is a topic for further investigation. Assume that at time $t = 0$ the external shear stress reaches the value necessary to overcome this obstacle. Then the dynamics of the transition (solution of system (6)) is described by the following formulae [38, 45]:

$$\frac{x}{t} = V(m), U = \frac{\varsigma - \varsigma^-}{\varsigma + \varsigma^-}, \Sigma_s = k = \frac{\pi(\varsigma + \varsigma^-)}{K\sqrt{m\varsigma\varsigma^-}}, W = kU$$

(8)

$$V = \frac{G - \varsigma^-}{G + \varsigma^-}, V^+ = \frac{1 - \varsigma^-}{1 + \varsigma^-}, V^- = \frac{G^- - \varsigma^-}{G^- + \varsigma^-},$$

(9)

where $G = \varsigma(E - K\sqrt{m_1})/(E + K\sqrt{m_1})$, $\varsigma = (1 - \sqrt{m_1})^2 / m$, $G^- \equiv G(m^-)$ and $\varsigma^- \equiv \varsigma(m^-)$, and $m^-$ is defined by the transcendental relation $\Sigma^- = \frac{\pi}{K(m^-)\sqrt{m^-}}$. The variable $V$ is the nonlinear group velocity of the wave solution in units of *c*. Along a line $x/t = V = $ constant in the *x-t* plane, all variables are constant. The solution is represented by a region expanding in time and bounded between the lines $x/t = V(m=0) = V^- = -1$ and $x/t = V(m=1) = V^+$. Note that inside the expanding region all variables are functions of time and position. The indices + and − designate the leading and trailing edges of the slip pulse, respectively. Thus the solutions (8 and 9) describe the dynamics of a slip pulse with two rupture fronts (or detachment fronts, in terms of equation (3)) propagating in opposite directions with different velocities $V^-$ and $V^+$. Figure 4 shows the ratio of $V^-$ to $V^+$ as a function of $V^-$. One can see that for small rupture velocities the ratio is large, i.e. the pulse propagates practically in one direction.

**Figure 3.** Schematics of spatial distribution of initial stress for solution of system (6) (solid line – analytical solution) and for system (7) (dotted line – computation solution).

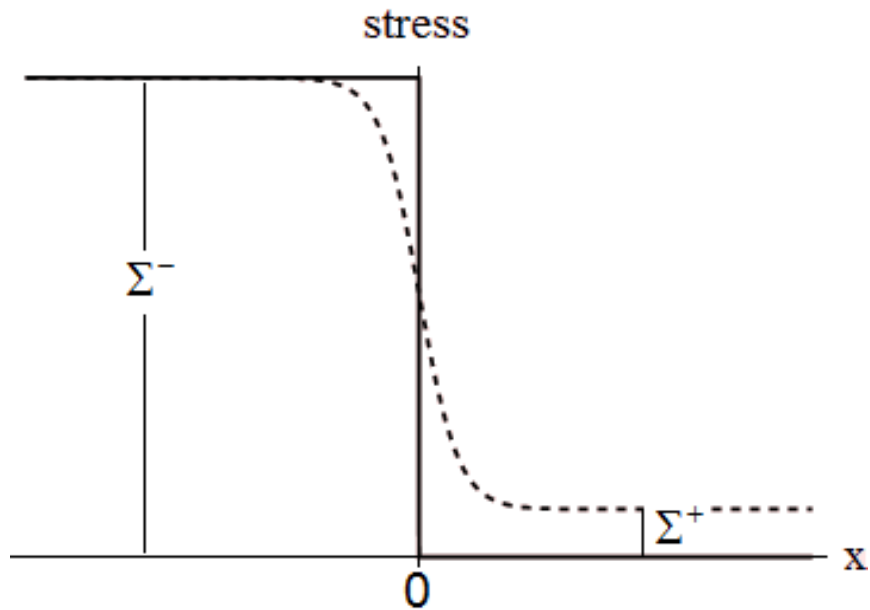

Figure 5 shows the result of the calculation of rupture velocity $V^-$ as a function of dimensionless initial shear stress $\Sigma^-$. Note a strong dependence of velocity on stress. Indeed, changes of shear stress by an order of magnitude lead to velocity changes by seven orders.

The velocity of dislocation movement $U(x,t)$ (formulae (8)) ranges from zero at the pulse trailing edge to the value $V^+$ at the pulse leading edge. The movement of dislocations is accompanied by slip with velocity $W(x,t)$ (formulae (8)). The slip velocity equals zero at the trailing and leading edges of a pulse and has a maximum value at $x=0$. Since $V(x=0)=0$ we can find the maximum slip velocity,

$$W(x=0) = \frac{\pi(\varsigma^0 - \varsigma^-)}{2K\sqrt{m^0 \varsigma^0 \varsigma^-}}, \qquad (10)$$

where $\varsigma^0 \equiv \varsigma(m^0)$ and $m^0$ is a solution of the equation $G(m^0) = \varsigma^-$. Figure 6 depicts the dependence of W on shear stress $\Sigma^-$. Note that in the case $\Sigma^- = \Sigma^+$ there is no transition process. For the chosen initial conditions $U(t=0,x)=0$, there is no slip, $W(t>0,x)=0$, and no rupture, $V^- = V^+ = 0$. If, however. $U(t=0,x)$ is a constant larger then 0, the solution is uniform sliding.

**Figure 4.** Ratio of fast rupture velocity $V^-$ to slow rupture velocity $V^+$ as a function of $V^-$. Both fast and slow velocities decrease with decreasing shear stress, but the latter velocity decreases more quickly. Thus, for processes such as slow slip events the slip pulse increases in extent essentially from one side.

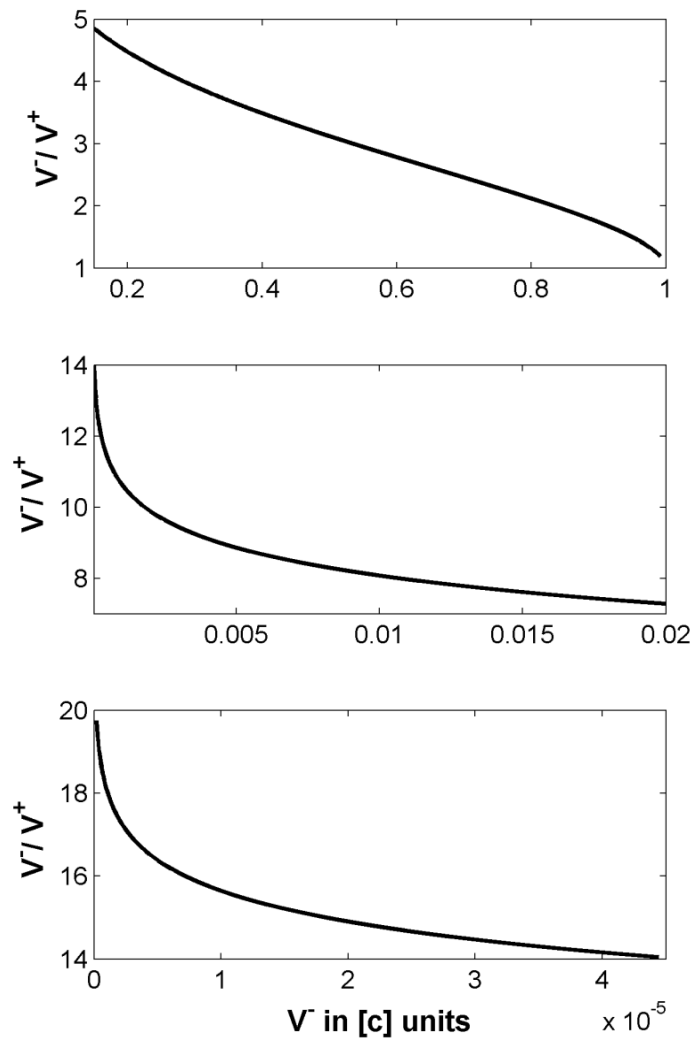

**Figure 5.** The dependence of rupture velocity on shear stress. Changes of shear stress by an order of magnitude lead to changes in $V^-$ by more than six orders of magnitude. The rupture velocity for the Parkfield earthquake (calculated based on observed data) and corresponding predicted shear stress are indicated by the dashed red lines on the top panel. The rupture velocity for an ETS event (calculated based on observed data) and predicted shear stress are indicated by the dashed green lines on the bottom panel. The predicted shear stress is used in Figure 6 to obtain the predicted slip velocity $W$.

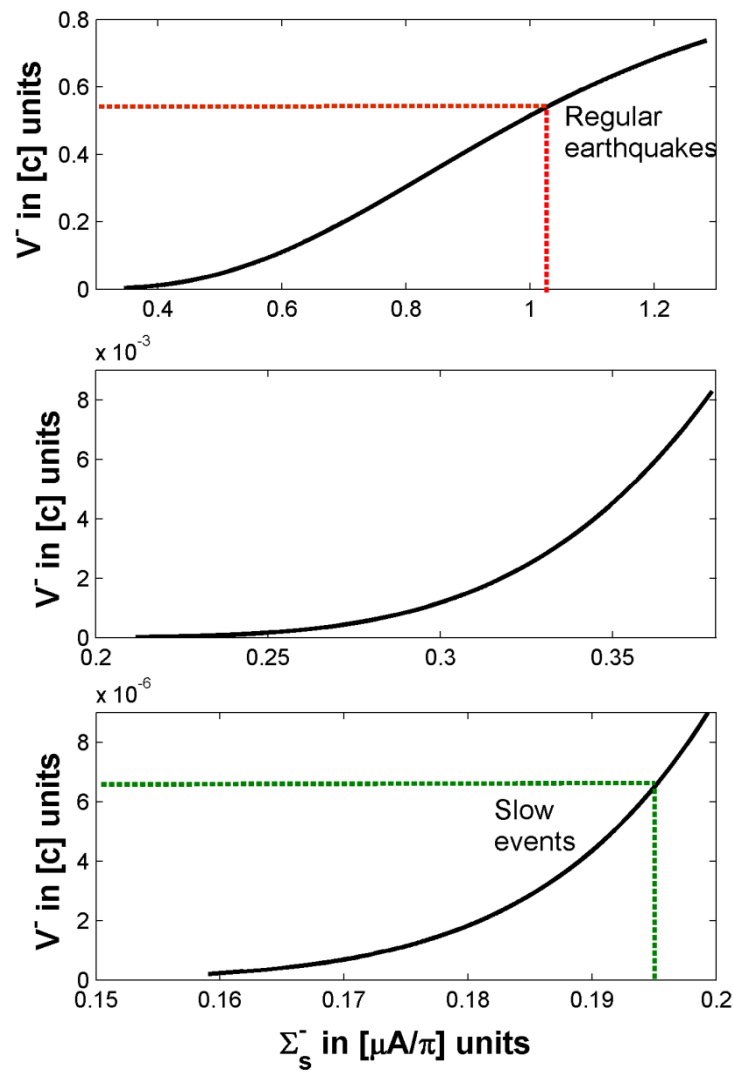

**Figure 6.** The dependence of maximum slip velocity *W* on shear stress. Changes of shear stress by an order of magnitude lead to changes in *W* by more than seven orders of magnitude. The predicted slip velocity for the Parkfield earthquake and shear stress (predicted based on observed rupture velocity (see Figure 5, top panel)) are indicated by the dashed red lines on the top panel. The predicted slip velocity for an ETS event and shear stress (predicted based on observed rupture velocity (see Figure 5, bottom panel)) are indicated by the dashed green lines on the bottom panel. Note that the value of shear stress for the Parkfied earthquake ($\Sigma^- \approx 1.03$) associated with the respective slip velocity corresponds to the value of shear stress calculated based on the rupture velocity (see Figure 5, top panel); however the value of shear stress for the ETS event ($\Sigma^- \approx 0.213$) is larger than the value predicted by the rupture velocity (see Figure 5, bottom panel). The reason for this discrepancy arises because dissipation was not taken into account (see next subsection).

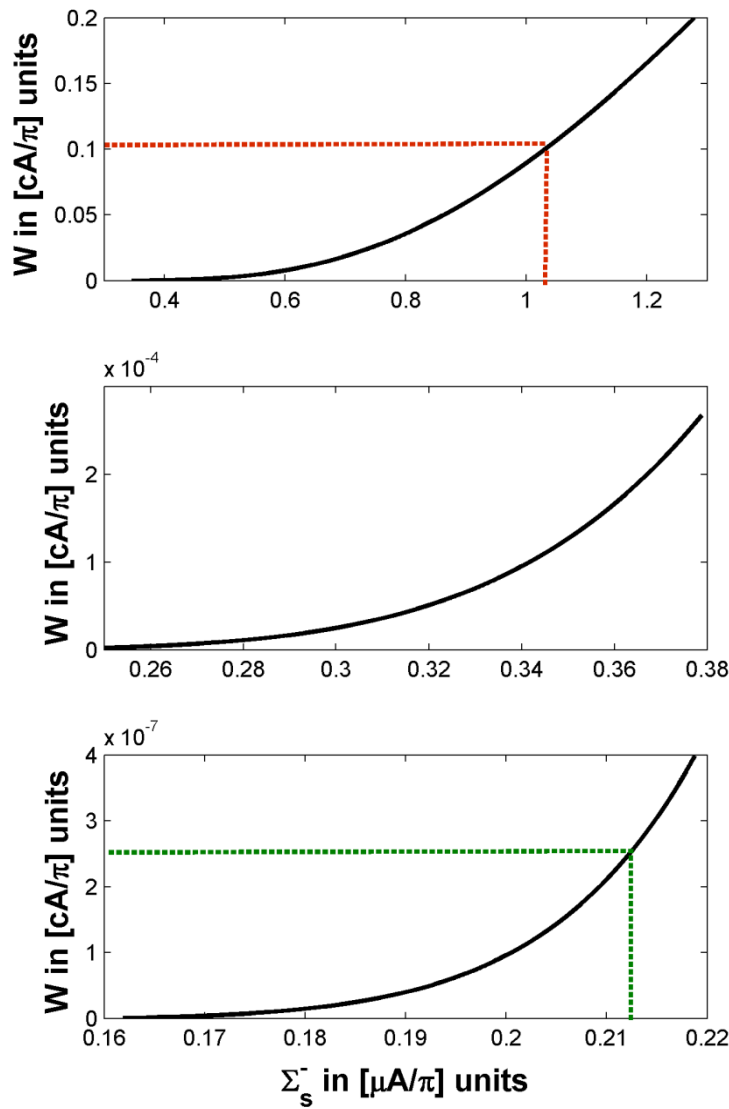

## 3.2. Solution with dissipation

Let us consider the same problem as above, i.e. the transition from the static to the dynamic regime, described now by equations (7). Note the small difference in initial conditions (dotted line at Figure 3) compared to the case in section 3.1 (solid line at Figure 3): 1) $\Sigma^+ \ll \Sigma^-$ but $\Sigma^+ > 0$ and 2) the stress distribution is a smooth function of *x* (not a step-function). These changes were necessary in order to obtain a numerically stable solution. For a simple velocity-dependent frictional term, $f = Cu_t$ and supposing that $F = 0$, we find $D(m,U) = CW(m,U)$, where *C* is the dissipation coefficient and *W* the slip velocity as defined by the formulae (8). Figure 7 depicts the spatial and temporal distribution of a slip pulse velocity for the various values of coefficient *C*. One can see that the amplitude of the slip velocity decreases with an increase in *C*.

**Figure 7.** Spatial and temporal distribution of slip pulse velocity for various values of coefficient *C*. The initial stresses, chosen arbitrarily to illustrate the effect of the dissipative term, are $\Sigma^- = 0.855$ and $\Sigma^+ = 0.179$. The length scale over which the initial stress

changes from $\Sigma^-$ to $\Sigma^+$ influences only the initial stage of slip pulse development. This influence becomes negligible when the two rupture fronts reach the area where $\Sigma_s = \Sigma^-$ for the trailing edge) and $\Sigma_s = \Sigma^+$ (for the leading edge).

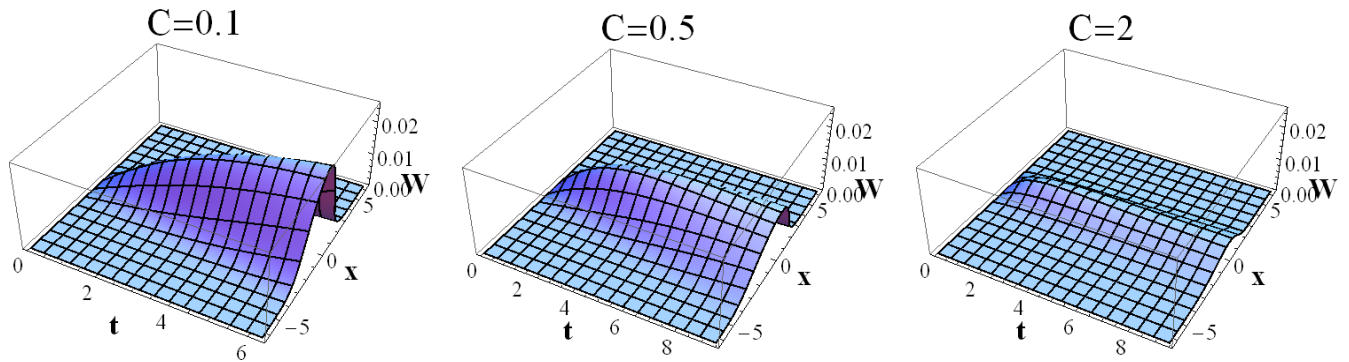

Figure 8 shows the slip velocities at time $2\pi$ for various values of *C*. To examine how friction (the dissipative term) affects the rupture velocity, we normalize *W* to make the maximum slip velocity the same for all cases. Then we see that both velocities $V^-$ and $V^+$ (left and right sides of slip pulse) are practically unchanged with increasing coefficient *C* for all cases considered, although the slip velocity amplitude changes by a factor greater than five. Note that this is true for any value of the initial stress $\Sigma^-$ provided $\Sigma^+ \ll \Sigma^-$. The spatial distribution of shear stress at time $2\pi$ is depicted in Figure 9. One can see that the larger dissipation, the smother is the transition from the stressed to the less stressed region. The result of calculating the dependence of slip velocity on the dissipative term for various initial shear stress values is shown in Figure 10. One can see that dissipation (dynamic friction) can essentially reduce the slip velocity. This velocity reduction depends weakly on the value of the initial shear stress $\Sigma^-$.

*3.3. Influence of initial stress $\Sigma^+$*

We have shown in section 3.1 how the rupture velocities and slip velocity depend on initial stress $\Sigma^-$. How do the $\Sigma^+$ and $\Delta\Sigma=\Sigma^--\Sigma^+$ values affect the shape and parameters of a slip pulse? Figure 11 depicts the spatial and temporal distribution of a slip pulse for various values of $\Sigma^+$. One can see that the shape of the pulse and its velocity do not change significantly unless the value of $\Sigma^+ \approx \Sigma^-$.. Analysis shows that $\Sigma^+$ does not affect the fast rupture velocity $V^-$; however the slow rupture velocity $V^+$ increases with increasing $\Sigma^+$ or decreasing $\Delta\Sigma$. When $\Delta\Sigma/\Sigma^- \ll 1$, $V^-$ is practically equal to $V^+$, i.e. the pulse is almost symmetric.

> **Figure 8.** Spatial distribution of slip pulse velocity, *W*, at $t = 2\pi$ for various values of the dissipation coefficient *C* in the friction relation $f = Cu_t$: *C*=0.1 (solid line), *C*=0.5 (dashed line) and *C*=2.0 (dotted line). The initial stresses are $\Sigma^- = 0.855$ and $\Sigma^+ = 0.179$. The amplitude of *W* has been multiplied by 1.8 for the dashed line and 5.2 for the dotted line to normalize to the same maximum slip velocity in all cases. The leading and trailing edges of the slip pulse (where $W \approx 0$) determine $V^-$ and $V^+$ (blue arrows), which are shown to vary little as a function of *C*, i.e., they are *independent of friction*. The bottom panel is obtained by plotting the top panel on a logarithmic scale.

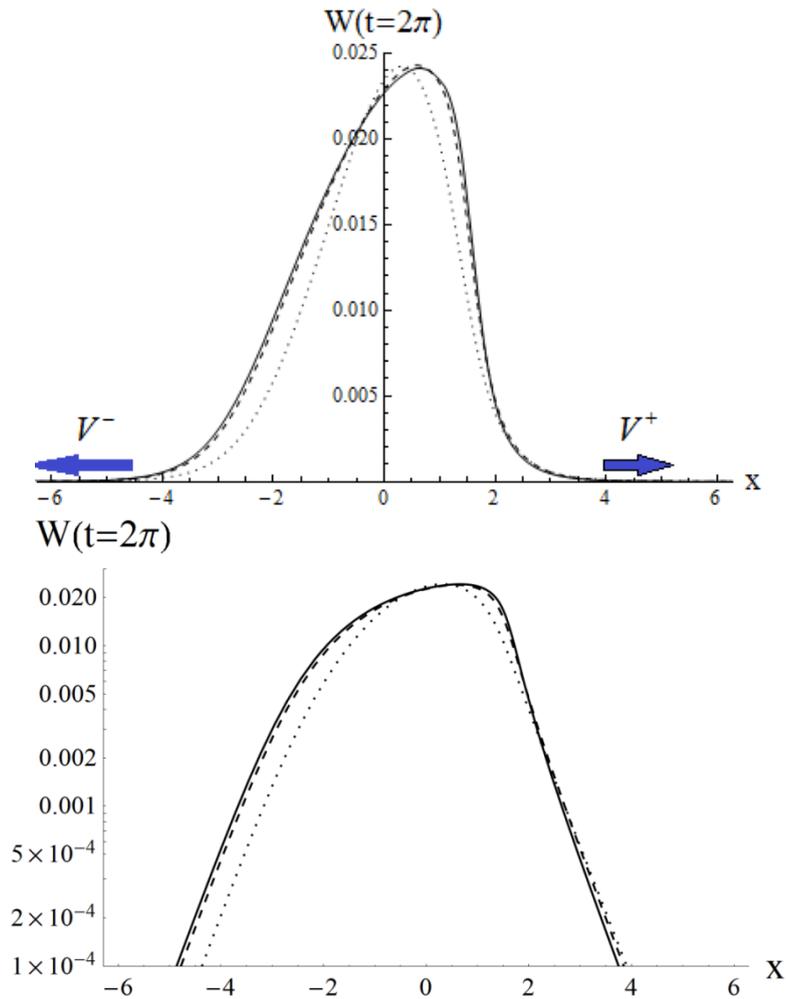

**Figure 9.** Spatial distribution of shear stress at $t = 2\pi$ for various values of the dissipation coefficient *C* in the friction relation $f = Cu_t$: *C*=0.1 (solid line), *C*=0.5 (dashed line) and *C*=2.0 (dotted line). The initial stresses are $\Sigma^- = 0.855$ and $\Sigma^+ = 0.179$.

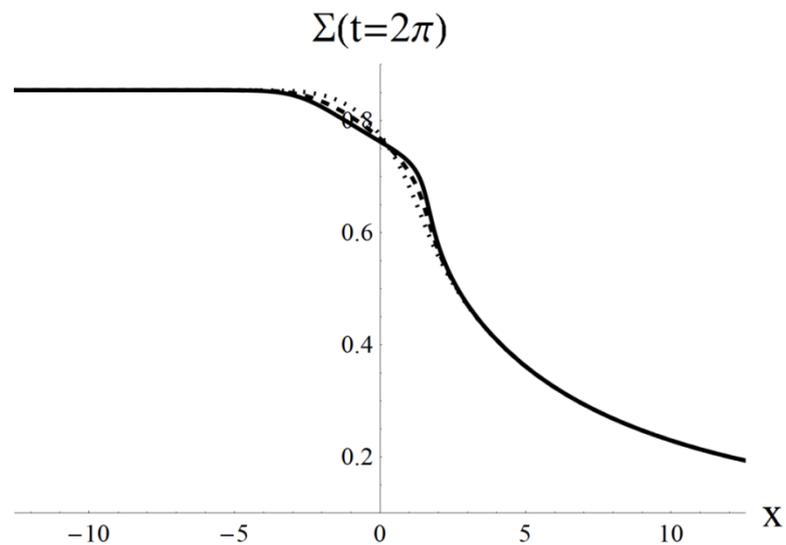

**Figure 10.** The ratio of slip velocity $W(C \geq 0)$ to slip velocity with no dissipation $W(C = 0)$ as function of dissipative coefficient C for various values of shear stress, $\Sigma^- = 0.855$ (solid curve), $\Sigma^- = 0.606$ (dashed curve) and $\Sigma^- = 0.419$ (dotted curve). Increasing friction significantly decreases the slip velocity, with a weak dependence on the initial shear stress $\Sigma^-$.

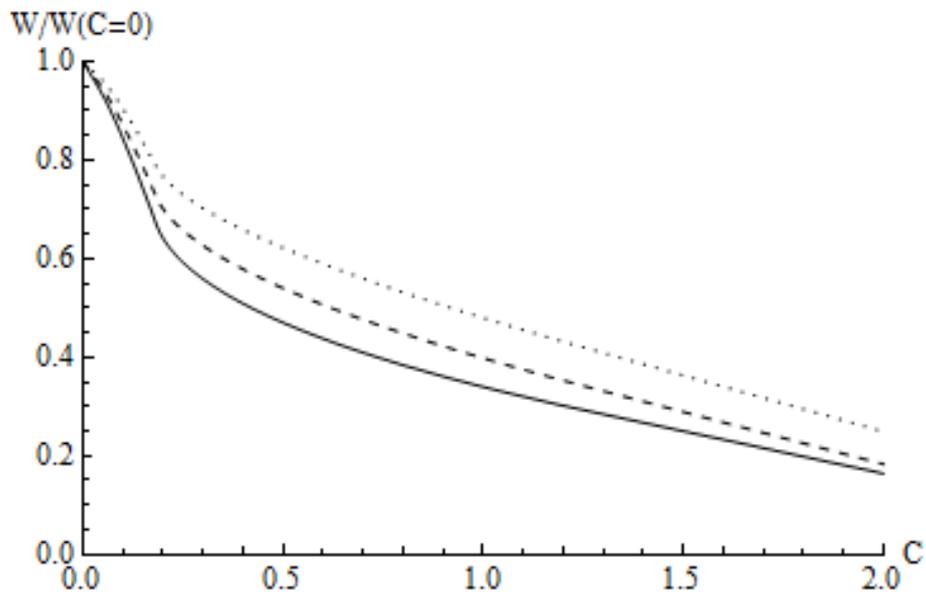

**Figure 11.** Spatial and temporal distribution of slip pulse velocity for various values of the coefficient $\Sigma^+$ (dissipation coefficient 0.1).

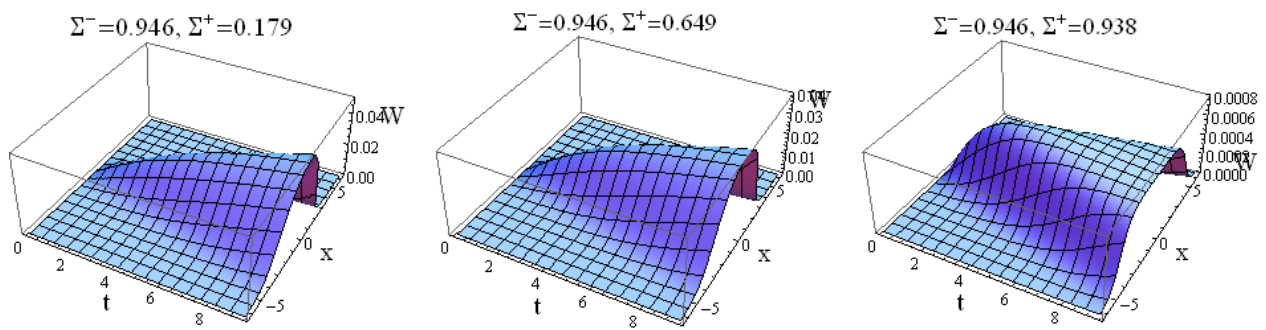

Figure 12 depicts the result of computing the slip velocity as a function of $\Delta\Sigma/\Sigma^-$ for two initial shear stress values $\Sigma^-$. Over a wide range of $\Delta\Sigma/\Sigma^-$ values, i.e. from 35% to 100%, the slip velocity is independent of $\Sigma^+$. However as $\Delta\Sigma/\Sigma^- \to 0$, the slip velocity also approaches zero.

**Figure 12.** Slip velocity at position $x = 0$ as a function of $\Delta\Sigma / \Sigma^-$ for two values of initial shear stress $\Sigma^-$.

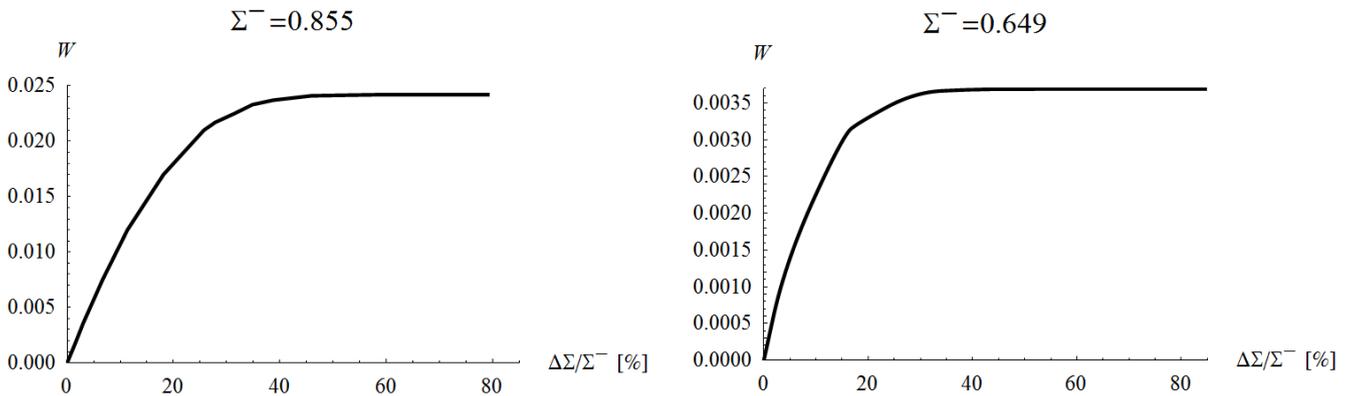

## 4. Discussion

Formulae (8-10) and Figures 4-6 connect the kinetic and dynamic parameters of the transition from static to dynamic friction. Using these formulae and figures one can find unknown parameters based on known parameters. Let us illustrate the applicability of our model by several examples.

Regular and slow earthquakes in the Earth's crust are typical transitional processes. Shear stress concentrated along a fault is relaxed due to various types of earthquakes and creep. Observations of far- and near-source ground motion allow reconstruction of kinematic parameters of regular earthquakes, i.e. rupture and slip velocities [47-51]. Dynamic parameters could be found by the modelling of processes using additional information and/or assumptions [51]. According to our model, the rupture velocity $V$ is explicitly defined by the initial stress $\Sigma^-$ and the elastic parameters of the medium, and doesn't depend on the dissipation term and the initial stress $\Sigma^+$ or stress difference $\Delta\Sigma$. Let us consider, as an example, the 2004 M=6 Parkfield earthquake. According to [51], the rupture velocity was about $V_{dim}^- = 3.0$ km/s and the slip velocity at the hypocenter area was about $W_{dim} = 0.5$ m/s. Taking the P-wave velocity in the Parkfield area to be $c_l = 6$ km/s and $v = 0.3$, the value of the parameter $c$ is 5.4 km/s, thus $V^- = V_{dim}^- / c \approx 0.55$. From Figure 5 (upper panel, red dashed lines) we find the dimensionless stress to be $\Sigma^- \approx 1.03$. Now we can find the dimensionless slip velocity (Figure 6, upper panel, red dashed lines) to be $W=0.1$. Recall that the velocity in Figure 6 is calculated for $C=0$. To find the velocity for other values of $C$ we need to use Figure 10. Thus, for the case considered, the dimensionless velocity is $W=0.083$, 0.05, and 0.02 for $C=0.1$, 0.5 and 2.0, respectively. Now we can calculate $A = \pi W_{dim} / (cW) \approx 0.0029$, 0.0035, 0.0058 and 0.0145 for the cases $C=0.0$, 0.1, 0.5 and 2.0, respectively. Supposing that $\mu = 30$ GPa we can calculate the initial stress to be $\Sigma_{dim}^- = \mu A \Sigma^- / \pi \approx 29$, 34, 58 and 145 MPa for these respective values of $C$. The result of a sophisticated dynamic modelling of the Parkfield earthquake [51] yields a stress value of about 31 MPa, which practically coincides with our estimate for the case of negligible dissipation. This leads us to the unexpected conclusion that (at least for some regular earthquakes) the dissipation due to friction may have practically no effect on the slip velocity, $W$, in addition to having no effect on the rupture velocities $V$ and $V^+$.

Now let us consider an example of slow earthquakes, e.g. so-called episodic tremor and slip phenomena [52-54]. It is known that during these events, a slip pulse slowly propagates along a subduction fault with an effective rupture velocity $V_{\text{dim}}^- \approx 10$ km/day ($V^- = V_{\text{dim}}^-/c \approx 2.14 \times 10^{-5}$) and with slip velocity $W_{\text{dim}} \approx 4mm/2\,\text{weeks}$ [55, 56]. From Figure 5 (lower panel, green dashed lines) we find a stress value $\Sigma^- \approx 0.212$. Then from Figure 6 (lower panel, green dashed lines) we find the dimensionless slip velocity to be $W \approx 2.6 \times 10^{-7}$. To estimate the actual value of *W* we need to know the value of *A*, which has been estimated by us in a previous work to be $A \approx 4 \times 10^{-5}$ [39]. Using this value we find the dimensionless velocity to be $W = \pi W_{\text{dim}}/(cA) \approx 0.5 \times 10^{-7}$. This value is one-fifth the value predicted by the model without dissipation. So for the slow event considered, friction has a pronounced influence on slip velocity. The corresponding dissipative coefficient (see Figure 10) is about *C*=2.0.

## 5. Conclusions

Our model for describing the transition from static to dynamic friction has been extended here to include dissipation. The significant and novel elements of the model are:
1) The FK model has been modified and adapted to the case of non-lubricant macroscopic friction by introducing a parameter *A* which is the ratio between the real and nominal area of the frictional surfaces;
2) In the model proposed, sliding occurs due to movement of a certain type of defect, a "macroscopic dislocation" (or area of localized stress), which requires much less shear stress than uniform displacement of frictional surfaces.
3) To describe measureable macroscopic parameters, we use the Whitham modulation equations applied to the SG equation rather than the SG equation itself which applies to details at the scale of individual dislocations.

The source of the transition from static to dynamic friction is the gradient of the shear stress (Figure 3). We show that in transition processes, the rupture velocity of the trailing edge *V*, i.e. the velocity of a detachment front propagating through the stressed material, is defined only by the initial stress $\Sigma^-$ and does not depend on friction in the case of a simple linear dissipative term considered here. This counterintuitive conclusion had already been predicted in our previous article [38], which neglects friction in the calculation, by comparing the accumulated elastic energy and dissipative energy released during the transition. Here we verified this prediction by explicitly including friction in the calculation (see Figures 8 and 11). Note that the initial stress (expressed in dimensional units) is $\Sigma_{\text{dim}}^- = \Sigma^- \mu A/\pi = \Sigma^- \mu \Sigma_N^-/(\pi\sigma_p)$, yielding an initial stress $\Sigma^- = (\pi\sigma_p/\mu)(\Sigma_{\text{dim}}^-/\Sigma_N)$, which is proportional to the ratio between shear and normal stress. Thus, the rupture velocity of the trailing edge *V* is defined by the ratio between shear and normal stress. This conclusion is consistent with the experimental result that the velocity of the detachment front is defined by this ratio [9]. The rupture velocity of the leading edge $V^+$, i.e. the velocity of the detachment front propagating from stressed to less stressed or unstressed material, is defined by the initial stresses $\Sigma^-$ and $\Sigma^+$ and likewise does not depend on friction. The velocity *V* is always larger than $V^+$ (Figure 4), i.e., the rupture front always propagates more easily through stressed rather than unstressed or less stressed material. However, in the case $(\Sigma^- - \Sigma^+)/\Sigma^- \ll 1$, these two velocities are almost equal, and hence the slip pulse is almost symmetric (see the right panel on Figure 11). Note that although the model formally does not include a rupture threshold, the value of $\Sigma^-$ may be considered as a threshold that is implicit in the model. As the threshold increases, the accumulated stress must increase in order to overcome this threshold. That is why regular earthquakes "require" larger stress $\Sigma^-$ than slow slip events.

Dynamic friction (dissipation) may reduce the slip velocity (Figures 7, 8 and 10). This effect is slightly larger in cases with larger initial shear stress $\Sigma^-$ (Figure 10). In some practical cases such as regular earthquakes, the influence of friction on slip velocity and pulse shape may be negligible. The effect of frictional processes (reduction of slip velocity) seems to be greater in slow slip events. In addition to initial shear stress and dissipation, slip velocity depends on $\Delta\Sigma = \Sigma^- - \Sigma^+$ (Figure 12). For small ratio $\Delta\Sigma/\Sigma^- \to 0$, the velocity $W \to 0$. However, the dependence of slip velocity on the shear stress gradient is negligible in the range $\Delta\Sigma/\Sigma^- > 0.4$.

The model developed here connects the kinetic and dynamic parameters of the transition process from static to dynamic friction. It allows description of the dynamics of the process under a very wide range of rupture and slip velocities, from velocities in regular earthquakes ($V^-/c \leq 1$, $W \approx 1$ m/s) down to the velocities in slow slip events ($V^-/c \approx 10^{-7}$, $W \approx 10^{-7}$ m/s). Both velocities depend critically on the accumulated stress, e.g., a change of stress by one order of magnitude can cause a velocity change by seven orders of magnitude (Figures 4 and 6).

We note that our model complements rate-and-state models and warrants further development to include elements of the latter, such as aging laws and dilatancy.

## Acknowledgments

This work was supported by NSF Grant EAR-1113578. We thank Norman Sleep and the two other (anonymous) reviewers for their thorough reading and insightful comments that significantly improved this paper.

## Author Contributions

Naum Gershenzon conceived the approach to the research, performed the calculations and analysis, and wrote the paper. Gust Bambakidis contributed to the analysis and writing of the manuscript. Thomas Skinner contributed to the calculations, analysis, and manuscript writing.

## Conflicts of Interest

The authors declare no conflict of interest.